\documentclass[aps,singlecolumn,amsmath,amssymb]{revtex4}

\usepackage{graphicx}
\usepackage{dcolumn}
\usepackage{bm}

\begin{document}

\title{Gamow-Teller transitions
  of neutron-rich $N=82,81$  nuclei by
  shell-model calculations}


\author{Noritaka Shimizu$^{1,*}$}
\author{Tomoaki Togashi$^{1}$}
\author{Yutaka Utsuno$^{2,1}$}
\affiliation{$^{1}$Center for Nuclear Study, The University of Tokyo,
  7-3-1 Hongo, Bunkyo-ku, Tokyo 113-0033, Japan \email{shimizu@cns.s.u-tokyo.ac.jp}}
\affiliation{$^{2}$Advanced Science Research Center,
  Japan Atomic Energy Agency, Tokai, Ibaraki 319-1195, Japan}



\begin{abstract}%
  $\beta$-decay half-lives of neutron-rich nuclei around $N=82$ are
  key data to understand the $r$-process nucleosynthesis.
  We performed large-scale shell-model calculations in this region 
  using 
  a newly constructed shell-model Hamiltonian, and successfully described 
  the low-lying spectra and half-lives of neutron-rich $N=82$ and $N=81$ isotones
  with $Z=42-49$ in a unified way. 
  We found that their Gamow-Teller strength distributions 
  have a peak in the low-excitation energies, which significantly 
  contributes to the half-lives. 
  This peak, dominated by $\nu 0g_{7/2} \to \pi 0g_{9/2}$ transitions,  
  is enhanced on the proton deficient side 
  because the Pauli-blocking effect caused by occupying 
  the valence proton $0g_{9/2}$ orbit is weakened. 
\end{abstract}

\maketitle

\section{Introduction}
\label{sec:intro}

The solar system abundances and their peak structures indicate that 
major origin of most elements heavier than iron is 
generated by the $r$-process nucleosynthesis
\cite{kajino-ppnp}.
A neutron-star merger was found by measuring the
gravitational wave which is followed by optical emission,
called ``kilonova'' \cite{nature-kilonova}.
The properties of neutron-rich nuclei are key issues 
to reveal the $r$-process nucleosynthesis which is expected
to occur in kilonova phenomena. 

The $r$-process path is considered to go through the neutron-rich
region of the nuclear chart. In the region where the $r$-process
path comes across the magic number $N=82$,
these nuclei form the waiting points of neutron captures in the $r$-process.
The path comes along the $N=82$ line in the chart bringing
about the so-called second peak
of the natural abundance formed by the astrophysical $r$-process nucleosynthesis.
In a typical $r$-process model, after reaching the $^{120}$Sr ($Z=38$, $N=82$),
the $\beta$-decay and the neutron capture are repeated alternately
to generate $N=82$ and $N=81$ nuclei 
up to $^{128}$Pd ($Z=46$, $N=82$) \cite{Shibagaki_APJ816}. 
This repeated process occurs if the $\beta$-decay rates of $N=81$ are smaller 
than their neutron-capture rates. 
Thus, the $\beta$-decay properties not only of the $N=82$ 
isotones but also of the $N=81$ ones 
are necessary to determine the $r$-process path, hence 
motivating the study of those very neutron-rich nuclei from 
the viewpoint of nuclear structure physics. 
Note that the properties of nuclei near $N=82$ 
are also awaited 
in the context of fission recycling 
\cite{fissionrecycling}.

On the experimental side,  
$\beta$-decay half-lives
of neutrino-rich nuclei around $N=82$ 
have recently been measured by the EURICA campaign 
conducted at the RI Beam Factory in RIKEN Nishina Center 
\cite{eurica,lorusso}.
More detailed data are now available for some nuclei. 
Many isomers have been identified near the $N=82$ shell gap, and 
some of their half-lives are obtained 
\cite{Cd128_130,watanabe_Pd128_Pd126,jangclaus_130Cd,manea_Cd132}. 
Furthermore, 
$\beta$-delayed neutron-emission probabilities and 
low-lying level structure have been measured 
\cite{Cd129-Saito,Wu_N82}.
These data 
provide a stringent test for nuclear-structure models. 
It should be noted that similar experimental activities are extended to 
the $N=126$ region, known as 
the third peak of the solar system abundance, for instance by
the KISS (KEK Isotope Separator System) project \cite{kekkiss}.

Many theoretical efforts have also been paid to 
systematically calculate $\beta$-decay half-lives 
such as by FRDM \cite{mollernix}, FRDM-QRPA \cite{FRDM-QRPA}, HFB-QRPA \cite{HFB-QRPA}, 
DFT-QRPA \cite{borzov_QRPA,DFT-QRPA}, and the gross theory \cite{gross}.
Recently, further sophisticated methods were introduced into the systematic
$\beta$-decay studies by introducing the FAM-QRPA \cite{Mustonen}
and by the relativistic CDFT-QRPA \cite{Martekin}.
Novel machine-learning techniques were also applied to
predict $\beta$-decay half-lives \cite{ML_beta}. 
The nuclear shell-model calculation is also one of
the most powerful theoretical schemes 
for this purpose. 
The previous shell-model studies are, however, restricted to calculating 
the half-lives of 
the singly-magic $N=82$
\cite{zhi,garcia,pinedo_prl83} 
and $N=126$ isotones \cite{zhi,suzuki_N126} 
due to the exponentially increasing dimensions of 
the Hamiltonian matrices in open-shell nuclei. 
The present work aims to extend those previous shell-model efforts 
to $N=81$ isotones within  
a unified description of the structures of neutron-rich
$N=82$ and $N=81$ isotones. 
The measured half-lives are well reproduced by the calculation, 
and we predict those for $^{125,126}$Ru, 
$^{124,125}$Tc and $^{124}$Mo. 
It is also predicted that these nuclei have rather strong GT 
strengths in the low excitation energies due to the increasing 
number of proton holes in the $g_{9/2}$ orbit, 
accelerating GT decay. 
This paper is organized as follows. 
The shell-model model space and its interaction are
defined in Sect.~\ref{sec:method}.
Section \ref{sec:smenergy} is devoted to 
the separation energies and low-lying spectra.
The Gamow-Teller strength distribution
and the half-lives are discussed in Sect.~\ref{sec:gt}.
Section \ref{sec:superGT} is devoted to the
discussion of the enhancement of the GT transitions
towards the proton-deficient nuclei and of its origin.
This paper is summarized in Sect.~\ref{sec:summary}.

\section{Framework of shell-model calculations}
\label{sec:method}

\begin{figure}[h]
  \centering
  \includegraphics[scale=0.6]{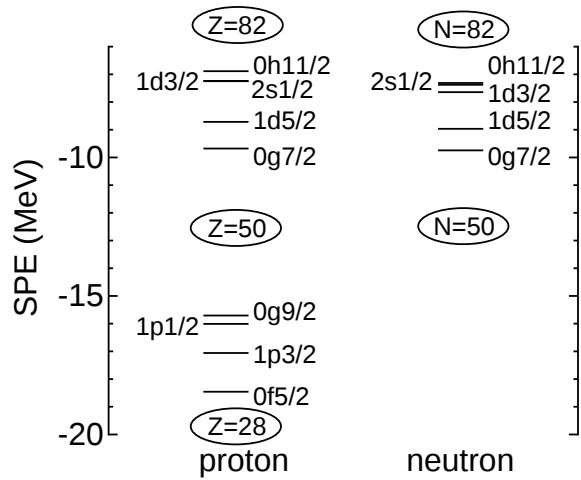}
  \caption{
    Single-particle energies for $^{132}$Sn
    determined from the experimental energy levels of its 
    one-particle and one-hole neighboring nuclei 
    \cite{Cd131-132,In131,AME2003,Sb133,Sn131-fogelberg,nudat2}.
    The single-particle orbits taken as the model space are shown.
  }
  \label{fig:spe}
\end{figure}

We performed large-scale shell-model calculations of $N=81$ and $N=82$
isotones.
The model space for the calculations
is taken as $0f_{5/2}$, $1p_{3/2}$, $1p_{1/2}$, $0g_{9/2}$, 
$0g_{7/2}$, $1d_{5/2}$, $1d_{3/2}$, $2s_{1/2}$, and $0h_{11/2}$ for 
the proton orbits
and $0g_{7/2}$, $1d_{5/2}$, $1d_{3/2}$, $2s_{1/2}$, and $0h_{11/2}$ for 
the neutrons orbits with the $^{78}$Ni inert core.
These orbits are shown in Fig.~\ref{fig:spe}.
Although we focus on $Z\le 50$ nuclei in this study, 
the single-particle orbits beyond the $Z=50$ shell gap are required
to be included in the model space explicitly so that
the Gamow-Teller transition causes the single-particle transition
of the valence neutrons beyond $N=50$ to the same orbits and its spin-orbit partners.
The model space is extended from 
that of the earlier shell-model study \cite{zhi}
by adding the proton $0f_{5/2}$, $1p_{3/2}$, and $0h_{11/2}$ orbits. 
In the preceding shell-model works ~\cite{zhi,garcia},
the proton $0h_{11/2}$ orbit was omitted 
to avoid the contamination of the spurious center-of-mass excitation,
although the neutron occupying $0h_{11/2}$ orbit can decay 
to the proton occupied in $0h_{11/2}$ by the Gamow-Teller transition.
In the present work,
we explicitly include the proton $0h_{11/2}$ orbits into the model space 
so that the proton single-particle orbits cover the whole neutron orbits.
For fully satisfying the Gamow-Teller sum rule the proton $0h_{9/2}$ orbit
is required, but its single-particle energy is too high to 
significantly affect the
Gamow-Teller strength of the low-lying states and it is omitted in the present work.
The contamination of spurious center-of-mass excitation
is removed by the Lawson method \cite{lawson} with $\beta_{CM} \hbar\omega/A=10$ MeV.
We truncate the model space by restricting up to 2 proton holes in $pf$ shell
and up to 3 protons occupying the orbitals beyond the $Z=50$ gap so that
the numerical calculation is feasible.
Even if applying such a truncation
the $M$-scheme dimension of the shell-model Hamiltonian matrix 
reaches $3.1\times 10^9$ and is quite large, 
and efficient usage of a supercomputer is essential.
The shell-model calculations were mainly performed on
CX400 supercomputer at Nagoya University and Oakforest-PACS at
The University of Tokyo and University of Tsukuba utilizing
the KSHELL shell-model code \cite{kshell}, which has been developed
for massively parallel computation.

The effective realistic interaction for the shell-model calculation
is constructed mainly
by combining the two established realistic interactions: 
the JUN45 interaction \cite{jun45} 
for the $f_5pg_9$ model space and the SNBG3 interaction \cite{snbg1}
for the neutron model space of $50<N,Z<82$.
The JUN45 and SNBG3 interactions were constructed from the G-matrix interaction
with phenomenological corrections using a chi-square fit to reproduce
experimental energies.
For the rest part of the two-body matrix elements (TBMEs), we adopt 
the monopole-based universal ($V_{MU}$) interaction \cite{vmu}
whose $T=1$ central force is scaled by the factor 0.75 
in the same way as Ref.~\cite{togashi-zr}.
The single particle energies are determined to reproduce the
experimental energies of one-nucleon neighboring nuclei of $^{132}$Sn
as shown in Fig.~\ref{fig:spe}. 
In addition, the strengths of the pairing interaction
and the diagonal TBMEs of the ($\pi 0g_{9/2}, \pi 0g_{9/2}$)
and ($\pi 0g_{9/2}, \nu 0h_{11/2}$) orbits are modified to
reproduce the experimental energy levels of $^{130}$Cd, $^{128}$Pd, and $^{130}$In.
The TBMEs are assumed to have the mass dependence $(A/132)^{-0.3}$.

\section{Separation energies and excitation energies}
\label{sec:smenergy}

\begin{figure}[t]
  \centering
  \includegraphics[scale=0.5]{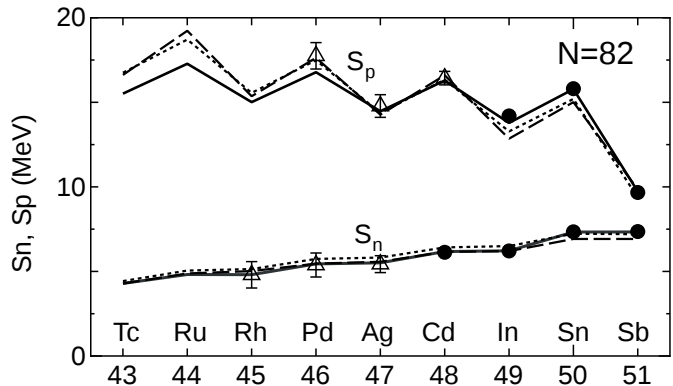}
  \caption{ Separation energies of $N=82$ isotones.
    The solid lines show 
    the proton and neutron separation energies provided
    by the present shell-model study.
    The filled circles and the open triangles with error bars
    denote the experimental values and 
    the extrapolated values from 
    the experimental systematics, respectively \cite{masstable}.
    The dotted lines and the dashed lines  are given by the KTUY mass formula \cite{ktuy05}
    and the FRDM \cite{mollernix}.
  }
  \label{fig:SnSp_N82}
\end{figure}

\begin{figure}[t]
  \centering
  \includegraphics[scale=0.5]{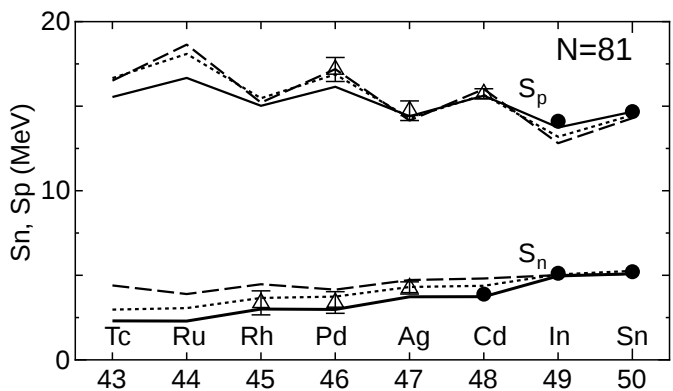}
  \caption{ Separation energies of $N=81$ isotones.
    See the caption of Fig.~\ref{fig:SnSp_N82} for details.
  }
  \label{fig:SnSp_N81}
\end{figure}

The binding energies and excitation energies of
the $N=82$ nuclei and those around them
are important not only for describing the $\beta$-decay properties, 
but also for confirming the validity of the shell-model interaction.
Figures \ref{fig:SnSp_N82} and \ref{fig:SnSp_N81} 
show the proton and neutron separation energies of the
$N=82$ and $N=81$ isotones, respectively.
The present shell-model results 
reproduce the experimental values excellently. 
The neutron separation energy determines
the threshold energy of the $\beta$-delayed
neutron emission, which is important for the $r$-process nucleosynthesis.
Since the $Q$-value of the $\beta^-$ decay is obtained using
the proton and neutron separation energies as 
\begin{eqnarray}
  Q(\beta^-, Z,N)
  &=& BE(Z+1,N-1)- BE(Z,N) + (m_n - m_p - m_e) c^2
      \nonumber \\
  &=& S_p(Z+1,N) - S_n(Z+1,N) + 0.782\ \textrm{MeV}, 
  \label{eq:qbeta}
\end{eqnarray}
where $BE(Z,N)$ denotes the binding energy of the $(Z,N)$ nucleus 
and 0.782 MeV is obtained from 
the mass difference of a neutron, 
a proton and an electron.
The $Q$ values of $\beta$-decay given 
by the shell-model results are in good agreement with the available
experimental values, as shown as the difference of
$S_n$ and $S_p$ in Figs.~\ref{fig:SnSp_N82} and \ref{fig:SnSp_N81}.
For comparison, the result of the KTUY \cite{ktuy05}
and the FRDM \cite{mollernix} mass formulae 
are also plotted in the figures, showing very good agreement with
the experimental values except slight underestimation in 
the proton separation energy of $^{130}$In. 
On the proton deficient side where the experimental values
are not available, the difference among the theoretical predictions  
gradually increases as the proton number
decreases, while the neutron separation energies of the $N=82$ isotones 
are rather close to one another.


\begin{figure}[t]
  \centering
  \includegraphics[scale=0.5]{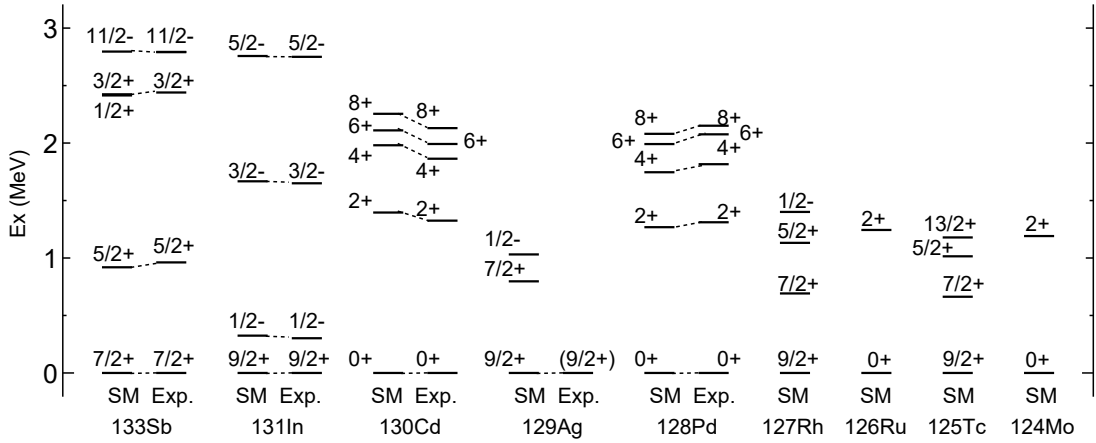}
  \caption{ Excitation energies of $N=82$ isotones:
    $^{133}$Sb, $^{131}$In, $^{130}$Cd, $^{129}$Ag, $^{128}$Pd,
    $^{127}$Rh, $^{126}$Ru, $^{125}$Tc, and $^{124}$Mo 
    compared between the shell model  
    (SM) and experiment (Exp.) \cite{nudat2}. }
  \label{fig:n82-ex}
\end{figure}

\begin{figure}[t]
  \centering
  \includegraphics[scale=0.55]{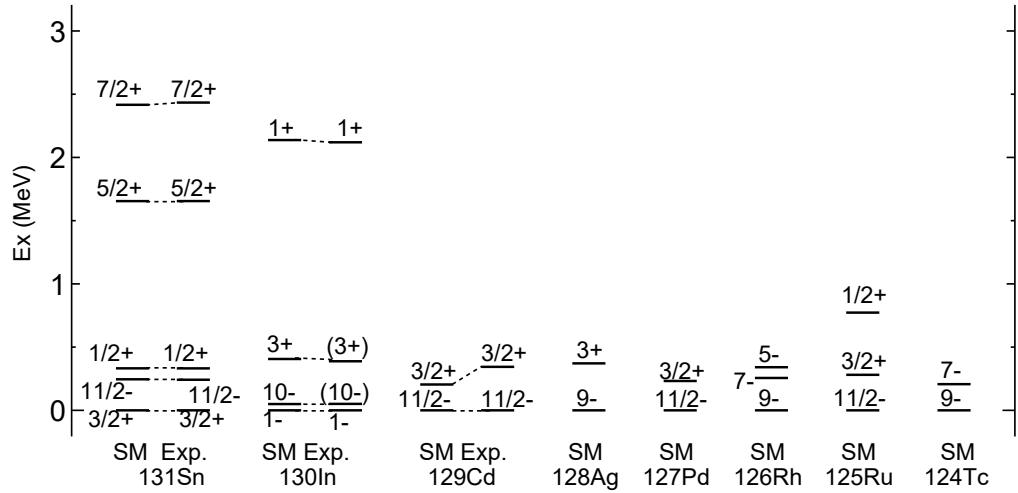}
  \caption{Excitation energies of $N=81$ isotones:
    $^{131}$Sn, $^{130}$In, and $^{129}$Cd,
    $^{128}$Ag, $^{127}$Pd,
    $^{126}$Rh, $^{125}$Ru, and $^{124}$Tc.
    See the caption of Fig.~\ref{fig:n82-ex} for details.
  }
  \label{fig:n81-ex}
\end{figure}

Figure \ref{fig:n82-ex}
shows low-lying energy levels 
in the neutron-rich $N=82$ isotones from $Z=42$
to $Z=50$. 
For the nuclei without data, we plot a few lowest levels 
obtained by the calculation. 
The calculated ground states are $0^+$ for the even-$Z$ isotopes 
and $9/2^+$ for the odd-$Z$ isotopes. 
The experimental levels are reproduced excellently 
by the shell-model results.
The levels of  $^{129}$Ag are experimentally unknown,
but two $\beta$-decaying states were found and tentatively assigned as
$9/2^+$ and $1/2^-$ \cite{nudat2} without their excitation energies known.  
In the present calculation, the $1/2^-$ state is located very close 
to the $7/2^+$ state. Considering a long $E3$ half-life in such a case, 
it is reasonable to assume that the $1/2^-$ state predominantly decays 
through $\beta$ emission. 

Figure \ref{fig:n81-ex} shows the 
excitation spectra of the $N=81$ isotones. 
Unlike the $N=82$ isotones, several candidates for 
the ground state and some $\beta$-decaying isomers are predicted. 
This is partly 
because the $1d_{3/2}$ and the $0h_{11/2}$ neutron orbits 
are located very close in energy as known in the spectra of $^{131}$Sn 
and the difference of their spin  
numbers is large. 
For $^{129}$Cd, two $\beta$-decaying states with $11/2^-$ and $3/2^+$ 
were known and their order had been controversial \cite{Cd128_130}.
A recent experiment concluded that 
its ground-state spin is $11/2^-$ 
and the excitation energy of $3/2^+$ is 343(8) keV \cite{manea_Cd132,Cd129-Saito},
which is consistent with our shell-model prediction.
For $^{127}$Pd, no experimental energy levels are known, and 
the present order of $11/2^-$ and $3/2^+$ agrees 
with another shell-model prediction \cite{watanabe_pd127}.
With regard to $\beta$-decay properties, 
the excitation energy of the
$1^+$ state of $^{130}$In plays a crucial role
in the $\beta$-decay half-life of $^{130}$Cd \cite{garcia}, 
whose $0^+$ ground state decays to the lowest $1^+$ state most strongly 
with the Gamow-Teller transition. 

\begin{figure}[t]
  \centering
  \includegraphics[scale=0.6]{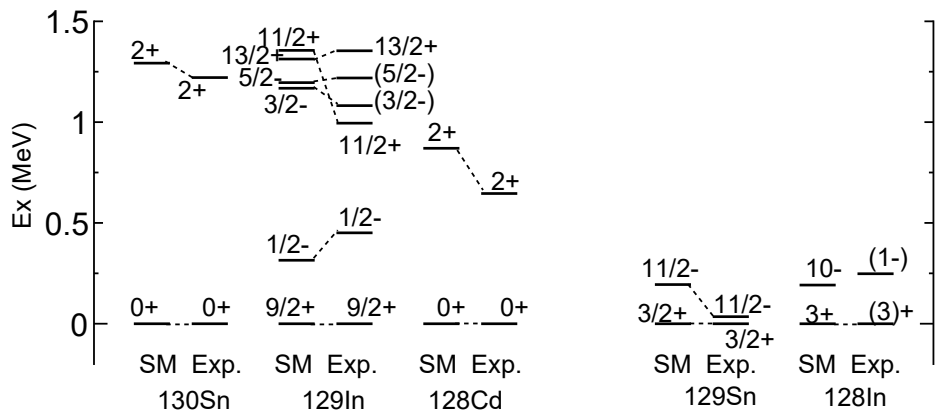}
  \caption{Excitation energies of the $N=80$ isotones ($^{130}$Sn, $^{129}$In, and $^{128}$Cd)
    and the $N=79$ isotones ($^{129}$Sn and $^{128}$In).
    The experimental values are taken from Refs.~\cite{nudat2,Cd129-Saito}.
    See the caption of Fig.~\ref{fig:n82-ex} for details.
  }
  \label{fig:n80-ex}
\end{figure}

Figure \ref{fig:n80-ex} shows the calculated energy levels 
of the $N=80$ and $N=79$ isotones for which the experimental 
data are available. 
We confirm a reasonable agreement between them.

The present calculation reproduces the 
experimental energies quite well,
thus confirming 
the validity of the model space and the effective interaction
employed in the present shell-model calculation.

\section{Gamow-Teller strength function and
  $\beta^-$-decay half-lives}
\label{sec:gt}

\begin{figure}[t]
  \centering
  \includegraphics[scale=0.42]{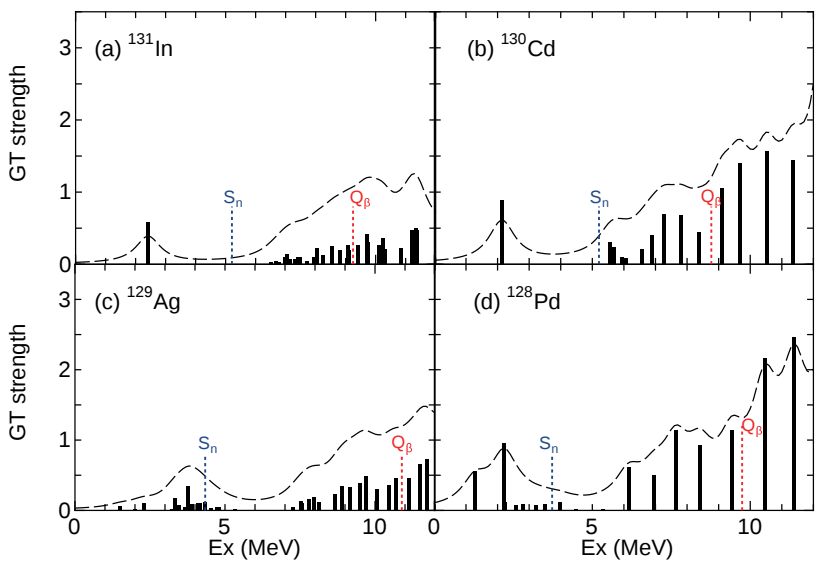}
  \caption{
    Gamow-Teller strength functions of $N=82$ isotones,  
    (a) $^{131}$In, (b) $^{130}$Cd, (c) $^{129}$Ag, 
    and (d) $^{128}$Pd against the excitation energies
    of the daughter nuclei.
    The dashed lines are the folded strength functions 
    by a Lorentzian function with the 1-MeV width.
    The values are shown without the quenching factor.
    The $Q_\beta$ values and the neutron separation energies 
    are shown as the red dotted lines and the blue dotted lines,
    respectively.
  }
  \label{fig:gt-N82}
\end{figure}
\begin{figure}[t]
  \centering
  \includegraphics[scale=0.42]{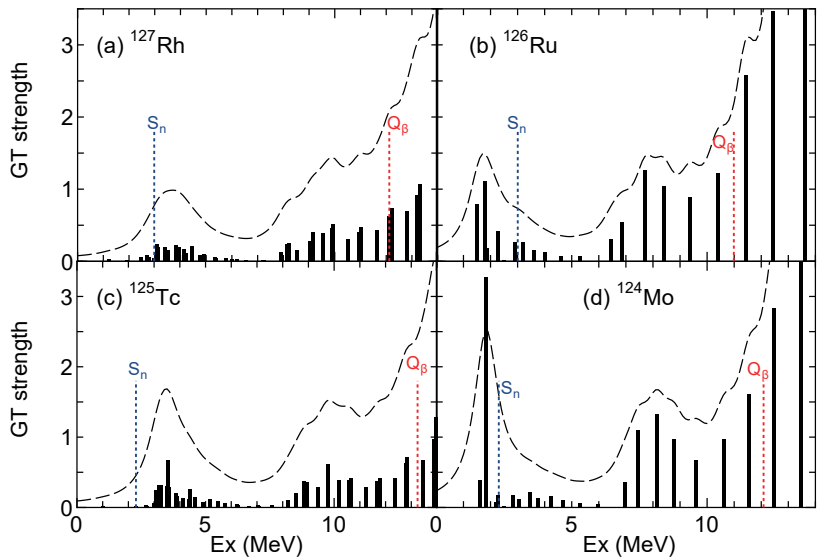}
  \caption{
    Gamow-Teller strength functions of proton-deficient $N=82$ isotones,  
    (a) $^{127}$Rh, (b) $^{126}$Ru, (c) $^{125}$Tc, 
    and (d) $^{124}$Mo.
    See the caption of Fig.~\ref{fig:gt-N82} for details.
  }
  \label{fig:gt-N82-2}
\end{figure}

We calculated the Gamow-Teller $\beta^-$-strength functions
for $N=82$ and $N=81$ neutron-rich nuclei 
to estimate their half-lives. 
We adopted the Lanczos strength function method
\cite{caurier_rmp,whitehead,utsuno_energy}
with 250 Lanczos iterations to obtain sufficiently converged 
results.
The magnitude of quenching of axial vector coupling is
still a challenging topic for nuclear
physics and has large uncertainty mainly caused by nuclear medium effect
and many-body correlations.
In the present work, the quenching factor 
is taken as $q_{\textrm{GT}}=0.7$,
which has been most widely used \cite{caurierPRL74,suzuki_N126}
and is consistent with
the adopted value of the preceding work, $q_{\textrm{GT}}=0.71$
\cite{garcia}.
The first-forbidden transition is omitted in the present work
because its contribution to the half-lives is small, around 13\%,  
and rather independent of nuclides
for the $Z=42-48$, $N=82$ isotones  
in a previous shell-model study \cite{zhi}. 
Furthermore, it is pointed out in \cite{Cd129-Saito} that 
a number of allowed transitions are observed in the $\beta^-$ decays 
of $^{121-131}$In and $^{121-125}$Cd, suggesting the dominance of 
GT transitions in the low excitation energies. 
This point will be discussed later.

Figure \ref{fig:gt-N82} shows the Gamow-Teller distributions of
$N=82$ isotones, $^{131}$In ($Z=49$), $^{130}$Cd ($Z=48$), 
$^{129}$Ag ($Z=47$), and $^{128}$Pd ($Z=46$).  
Figure \ref{fig:gt-N82-2} shows those of
more proton-deficient $N=82$ isotones, 
$^{127}$Rh ($Z=45$), $^{126}$Ru ($Z=44$), 
$^{125}$Tc ($Z=43$), and $^{124}$Mo ($Z=42$).
The $Q$-values are taken from the experiments for $^{131}$In and $^{130}$Cd
\cite{nudat2}, while the present theoretical
$Q$-values are used for the other nuclei.
These figures present a very remarkable systematics of low-energy 
GT strength distributions, 
which play a crucial role in those $\beta$-decay half-lives. 
First, all the $N=82$ isotones considered here have strong GT strengths 
in the low-excitation energies. Except $^{131}$In,
they are peaked at $\sim 3.5$~MeV and $\sim 2$~MeV 
for the odd-$Z$ and even-$Z$ parents, respectively, 
and the GT strengths are more 
concentrated for the even-$Z$ isotopes. This odd-even effect 
is in accordance with what is found in the $sd$-$pf$ shell 
region \cite{syoshida_PRC97}. 
Second, this low-energy GT peak grows with decreasing 
proton number. 
This is an interesting feature of low-energy Gamow-Teller transitions 
predicted for this region, and more detailed discussions will be given in Sec.~\ref{sec:superGT}. 



\begin{table}
  \centering
  \begin{tabular}{l||c|c|c|c|c|c|c}
    $T_{1/2}$(ms), $N=82$              & SM$^{\textrm{th}}$& SM$^{\textrm{exp}}$ & SM13   & SM07 & SM99   & Exp15 & Exp16  \\
    \hline
    $^{131}$In $\rightarrow$ $^{131}$Sn & 156  &   154  & 247.53 & 260  & 177   & 261(3) & 265(8)\\
    $^{130}$Cd $\rightarrow$ $^{130}$In & 158  &   116  & 164.29 & 162  & 146   & 127(2) & 126(4)\\
    $^{129}$Ag $\rightarrow$ $^{129}$Cd &  44  &        & 69.81 &  70  &  35.1 &  52(4) & \\
    $^{128}$Pd $\rightarrow$ $^{128}$Ag &  28  &        & 47.25 &  46  &  27.3 &  35(3) & \\
    $^{127}$Rh $\rightarrow$ $^{127}$Pd & 13.9 &        & 27.98 &27.65 &11.8 &  20$^{+20}_{-7}$ &\\
    $^{126}$Ru $\rightarrow$ $^{126}$Rh &  9.2 &        & 20.33 &19.76 & 9.6 &        & \\
    $^{125}$Tc $\rightarrow$ $^{125}$Ru &  5.7 &        & 9.52 & 9.44 &  4.3 & \\
    $^{124}$Mo $\rightarrow$ $^{124}$Tc &  4.0 &        & 6.21 & 6.13 &  3.5 & \\
  \end{tabular}
  \caption{$\beta$-decay half-lives of the $N=82$ isotones
    by the present shell-model calculations (SM),
    the shell-model study with the experimental $Q$ value (SM$^{\textrm{exp}}$),
    the earlier shell-model works (SM13 \cite{zhi}), (SM07 \cite{garcia}),
    (SM99 \cite{pinedo_prl83}),
    and the recent experiments (Exp15) \cite{lorusso}, (Exp16) \cite{Cd128_130}.
    The half-lives are shown in the unit of ms.
    \label{tab:N82life}
  }
\end{table}

Table \ref{tab:N82life} shows 
the $\beta$-decay half-lives of the $N=82$ isotones.
The half-life is estimated by accumulating the transition
probabilities from the parent ground state to the daughter states
whose excitation energies are below the $Q_\beta$ value.
The shell-model results show reasonable agreement 
with the experimental values.
While the present half-lives of $^{129}$Ag and $^{128}$Pd
are closer to the experimental values than the earlier shell-model result, 
the half-life of $^{131}$In is underestimated.
This underestimation is caused by the large GT transition
to the lowest $7/2^+$ state of
the daughter $^{131}$Sn at $E_x=2.4$ MeV, which might imply
the need for further improvement of the theoretical model.
This state is considered to be dominated by the $\nu 0g_{7/2}$-hole
state of $^{132}$Sn. In the pure $\pi 0g_{9/2}^{-1} \to \nu 0g_{7/2}^{-1}$ 
single-particle transition, the corresponding $B(\textrm{GT})$ value 
is as much as 1.78 
without the quenching factor introduced. 
On the other hand, the present calculation gives $B(\textrm{GT})=0.58$. 
This value is considerably reduced from the single-particle value 
due to configuration mixing,  
but further reduction is required to completely 
reproduce the data.

For comparison, Table \ref{tab:N82life} also shows 
three shell-model results by the Strasbourg group:
SM13 \cite{zhi},  SM07 \cite{garcia},
and SM99 \cite{pinedo_prl83}.
The half-lives of $^{126}$Ru, $^{125}$Tc, and $^{124}$Mo
predicted by the present calculation are close to those of
SM99 \cite{pinedo_prl83}.
The half-lives of SM13 \cite{zhi} and SM07 \cite{garcia}
are quite close to each other.
While the first-forbidden transition was omitted
and the quenching factor of the Gamow-Teller transition
was taken as $q_{\textrm{GT}}=0.71$ in SM07, 
the first-forbidden transition is included
with $q_{\textrm{GT}}=0.66$ in SM13.
The agreement of these results indicates
that the contribution of the first-forbidden decay
is rather independent of the nuclides and can be
absorbed into the minor change of the Gamow-Teller
quenching factor in this mass region.

\begin{figure}[t]
  \centering
  \includegraphics[scale=0.5]{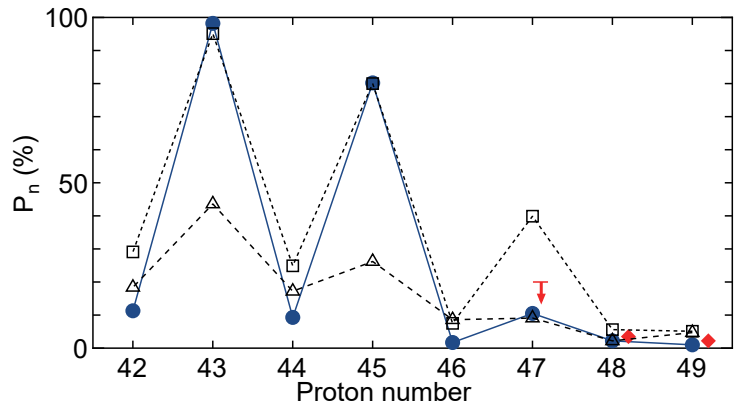}
  \caption{
    Neutron emission probabilities of $N=82$ isotones.
    The blue filled circles, black open squares, and black open triangles
    denote the results by the present work, 
    the earlier shell-model work \cite{zhi}, 
    and the FRDM+QRPA \cite{FRDM-QRPA}, respectively.
    The red diamond denotes the experimental value 
    and the red line with an arrow at $Z=47$ denotes the experimental
    upper limit \cite{Pn_Cd130,NDS_beta}.
  }
  \label{fig:Pn-N82}
\end{figure}

$\beta$-delayed neutron emission is important 
for understanding the freezeout of the $r$ process \cite{kajino-ppnp}.
Figure  \ref{fig:Pn-N82} show
$\beta$-delayed neutron emission probabilities $P_n$ for 
$N=82$ nuclei. 
In the present calculation, we accumulate the probabilities of
the $\beta$-decay
to the states above the neutron-emission threshold $S_n$ to
obtain $P_n$.
The present shell-model results show an odd-even staggering
similar to that of the earlier shell model \cite{zhi}, 
while the FRDM-QRPA results show weaker odd-even staggering. 
This odd-even staggering is caused by 
the difference of the peak position 
and the degree of concentration of the Gamow-Teller transition strengths. 
As discussed already using Figs.~\ref{fig:gt-N82} and \ref{fig:gt-N82-2}, 
the GT peaks of the even-$Z$ parent nuclei 
are located at around $E_x=2$ MeV, 
which is lower than $S_n$, causing their small $P_n$ values. 
For $^{124}$Mo, it is predicted that this low-energy GT 
strength is concentrated by 
a single peak that is located slightly below $S_n$. Hence its 
$P_n$ is very sensitive to the detail of the energies concerned. 
For the odd-$Z$ nuclei of $^{127}$Rh and $^{125}$Tc, 
the low-energy GT peak is located higher than $S_n$, enlarging 
their $P_n$ values.

\begin{figure}[htbp]
  \centering
  \includegraphics[scale=0.42]{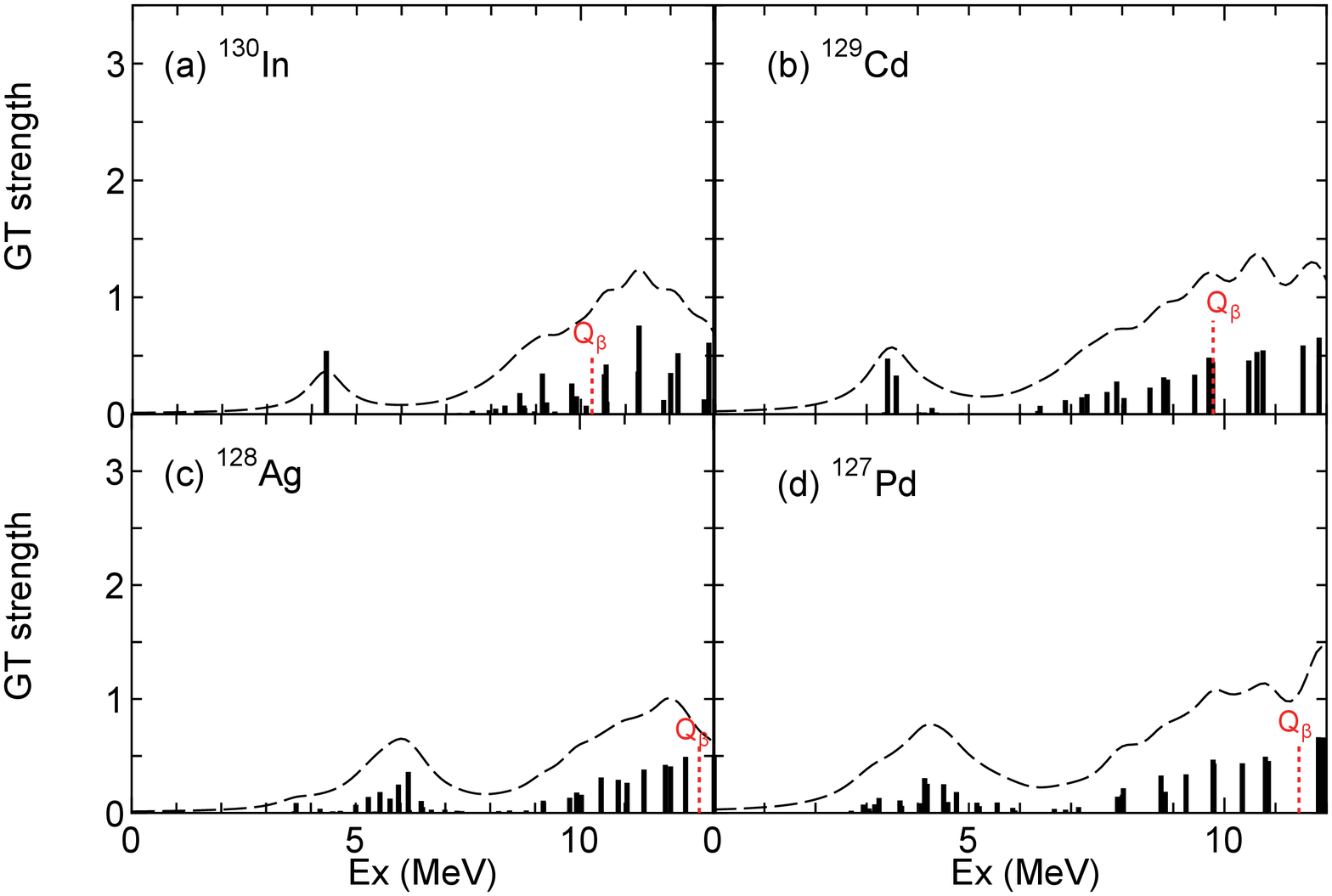}
  \caption{
    Gamow-Teller strength functions of $N=81$ isotones,  
    (a) $^{130}$In, (b) $^{129}$Cd, (c) $^{128}$Ag,
    and (d) $^{127}$Pd.
    See the caption of Fig.~\ref{fig:gt-N82} for details.
  }
  \label{fig:gt-N81}
\end{figure}

\begin{figure}[htbp]
  \centering
  \includegraphics[scale=0.42]{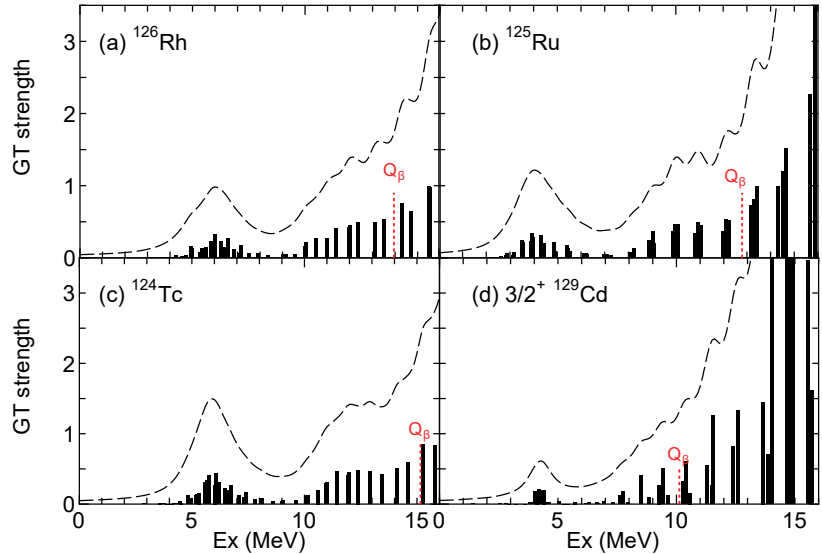}
  \caption{
    Gamow-Teller strength functions of $N=81$ isotones,  
    (a) $^{126}$Rh, (b) $^{125}$Ru, (c) $^{124}$Tc,
    and (d) the isomeric $3/2^+$ state of $^{129}$Cd.
    See the caption of Fig.~\ref{fig:gt-N82} for details.
  }
  \label{fig:gt-N81-2}
\end{figure}

Figures  \ref{fig:gt-N81} and \ref{fig:gt-N81-2} 
show the Gamow-Teller $\beta^-$-strength distribution
of $N=81$ isotones, namely $^{130}$In, $^{129}$Cd, $^{128}$Ag,
$^{127}$Pd, $^{126}$Rh, $^{125}$Ru and $^{124}$Tc 
obtained by the present shell-model
calculations.
Figures  \ref{fig:gt-N81-2} shows also 
the distribution of the isomeric $3/2^+$ state of $^{129}$Cd.
The $Q(\beta^-)$ values are taken from experiments for $^{130}$In
and $^{129}$Cd \cite{nudat2},
and taken from shell-model values
for the other nuclei. The low-energy GT peaks are obtained in 
all the cases calculated. They are located higher for the 
odd-$Z$ parents due to pairing correlation in the daughter nuclei, 
but fragmented in a similar manner. 
Like the case of the $N=82$ isotones, those peaks are enhanced 
as the proton number decreases and 
the proton $0g_{9/2}$ orbit becomes unoccupied.


Table \ref{tab:N81life} shows 
the $\beta$-decay half-lives of the $N=81$  isotones.
The half-lives of the five nuclei with $Z\ge 45$ 
show reasonable agreement with the available experimental values, 
indicating the validity of the present shell-model calculation.
The half-life of the $3/2^+$ isomeric state of $^{129}$Cd 
is also shown in the table to demonstrate 
the capability to obtain the $\beta$-decay rates of isomeric states.

\begin{table}
  \centering
  \begin{tabular}{l||c|c|c|c}
    $T_{1/2}$(ms), $N=81$  & SM$^{\textrm{th}}$ &SM$^{\textrm{exp}}$  & Exp15  &  Exp16 \\
    \hline
    $^{130}$In $\rightarrow$ $^{130}$Sn & 286 & 311 & 284(10) \\
    $^{129}$Cd $\rightarrow$ $^{129}$In & 182 & 139 & 154.5(20) & 147(3) \\
    $^{129}$Cd ($\frac32^+$) $\rightarrow$ $^{129}$In & 266 & 181  &  & 157(8) \\
    $^{128}$Ag $\rightarrow$ $^{128}$Cd &  49 & &  59(5) \\
    $^{127}$Pd $\rightarrow$ $^{127}$Ag &  32 & &  38(2) \\
    $^{126}$Rh $\rightarrow$ $^{126}$Pd &  17 & &  19(3) \\
    $^{125}$Ru $\rightarrow$ $^{125}$Rh &  11 & &        \\
    $^{124}$Tc $\rightarrow$ $^{124}$Ru &  7.0& &        \\
  \end{tabular}
  \caption{$\beta$-decay half-lives of the $N=81$ isotones
    obtained by the present shell-model study (SM$^{\textrm{th}}$),
    the shell-model study with the experimental $Q$ value (SM$^{\textrm{exp}}$),
    and the experiments (Exp15) \cite{lorusso}, 
    (Exp16) \cite{Cd128_130}.
    The half-lives are shown in the unit of ms.
    \label{tab:N81life}
    The half-life of the $3/2^+$ isomeric state of $^{129}$Cd is
    also shown.
  }
\end{table}

In Tables \ref{tab:N82life} and \ref{tab:N81life},
SM$^{\textrm{th}}$ and SM$^{\textrm{exp}}$
show the shell-model results using the shell-model $Q$ value
and those using the experimental $Q$ value, respectively,
to discuss the uncertainty of
the present theoretical model.
The deviations of the choice of the $Q$ values show up to 30\% at most.
The fitted quenching factor to reproduce the experimentally measured
half-lives of $^{129}$Cd, $^{130}$Cd and the 3/2$^+$ isomer by
the SM$^{\textrm{exp}}$ result 
is $q_{\textrm{GT}}=0.67$, which shows a 9\% increase of the
half-life estimate. These differences are considered
as the uncertainties of the present model.

\section{Possible occurrence of superallowed Gamow-Teller transitions toward $Z=40$}
\label{sec:superGT}

\begin{figure}[t]
  \centering
  \includegraphics[scale=0.5]{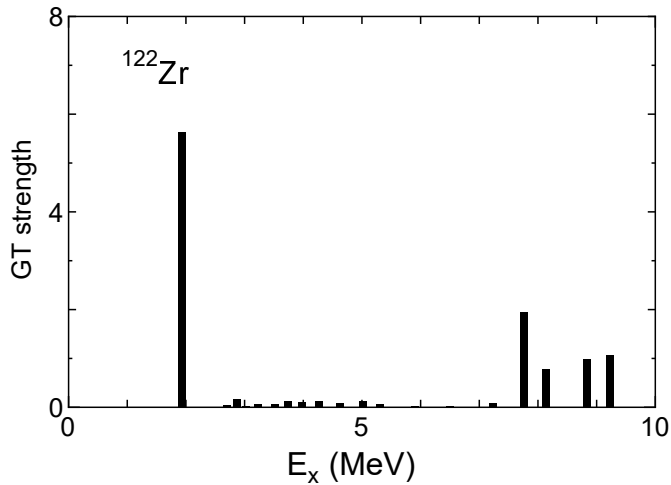}
  \caption{ Gamow-Teller strength functions of $^{122}$Zr. See the 
caption of Fig.~\ref{fig:gt-N82} for details. 
  }
  \label{fig:zr122}
\end{figure}

As mentioned in the last section, 
Figures~\ref{fig:gt-N82} and \ref{fig:gt-N82-2} show that for the 
even-$Z$ parents a low-energy Gamow-Teller peak emerges at $\sim 2$~MeV and 
that its magnitude is enhanced as the proton number decreases. 
As depicted in Fig.~\ref{fig:zr122}, this 
peak is finally concentrated in a single state 
at $^{122}$Zr with $Z=40$, leading to $B(\textrm{GT})=2.7$ calculated 
with the quenching factor $0.7$. 
In this section, we focus on this growing Gamow-Teller peak 
toward $Z=40$. 

At first, we discuss why this peak is enlarged with decreasing $Z$. 
By analyzing one-body transition densities obtained in the present calculations, 
one can see that those low-energy Gamow-Teller peaks 
are dominated by the $\nu 0g_{7/2} \to \pi 0g_{9/2}$ transition. 
If the $\pi 0g_{9/2}$ orbit is completely filled, this transition 
does not occur due to the Pauli blocking. This 
blocking effect is weakened by removing protons from the $\pi 0g_{9/2}$ orbit, 
hence the enlargement of the low-energy Gamow-Teller peak. 

The resulting $B(\textrm{GT})$ values of this peak are particularly large 
at $^{124}$Mo and $^{122}$Zr compared to typical values. 
It is known from the systematics \cite{NDS_logft} that the $\log ft$ values 
of allowed $\beta$ decays are distributed around 
$\log ft\sim 6$, which 
corresponds to $B(\textrm{GT})\sim 10^{-3}$-$10^{-2}$ for Gamow-Teller transitions. 
A well-known deviation from this systematics is the superallowed 
(Fermi) transition. When isospin is a good quantum number, the Fermi transition 
occurs only between isobaric analog states, giving a typical $\log ft$ 
of 3.5. With regard to Gamow-Teller transitions, however, there are only a few 
cases where the $\log ft$ value is comparable to those of the superallowed Fermi transitions 
because of the fragmentation of Gamow-Teller strengths. 
Since $B(\textrm{GT})=1$ leads to $\log ft = 3.58$, 
the $B(\textrm{GT})$ value of the order of unity  
is a good criterion to compare the superallowed Fermi transition. 

It is proposed in \cite{brownPPNP01} that such extraordinarily fast Gamow-Teller 
transitions be classified as Super Gamow-Teller transitions. 
At that time, only two Gamow-Teller transitions, $^{6}$He$\to ^{6}$Li and 
$^{18}$Ne$\to ^{18}$F, were known to satisfy the condition 
of Super Gamow-Teller transition defined in \cite{brownPPNP01}, i.e., 
$B(\textrm{GT}) > 3$. 
These large Gamow-Teller strengths are caused by the constructive interference 
of $j_> \to j_>$ and $j_> \to j_<$ matrix elements \cite{fujitaEPJA20}. 
It was also predicted in \cite{brownPPNP01} that two $N=Z$ doubly-magic nuclei 
$^{56}$Ni and $^{100}$Sn were candidates for nuclei causing Super Gamow-Teller 
transitions. 
Although the Gamow-Teller strengths from $^{56}$Ni were measured to be fragmented 
about a decade later \cite{sasanoPRL11}, $^{100}$Sn is now established to 
have a very large $B(\textrm{GT})$ value ($9.1^{+3.0}_{-2.6}$ in \cite{hinke_Nature12} 
or $4.4^{+0.9}_{-0.7}$ in \cite{lubosPRL19}) 
to a $1^+$ state located at around 3~MeV.  
This Gamow-Teller decay is called ``superallowed Gamow-Teller'' decay in \cite{hinke_Nature12} 
on the analogy of the superallowed Fermi decay. 

\begin{figure}[t]
  \centering
  \includegraphics[scale=0.4]{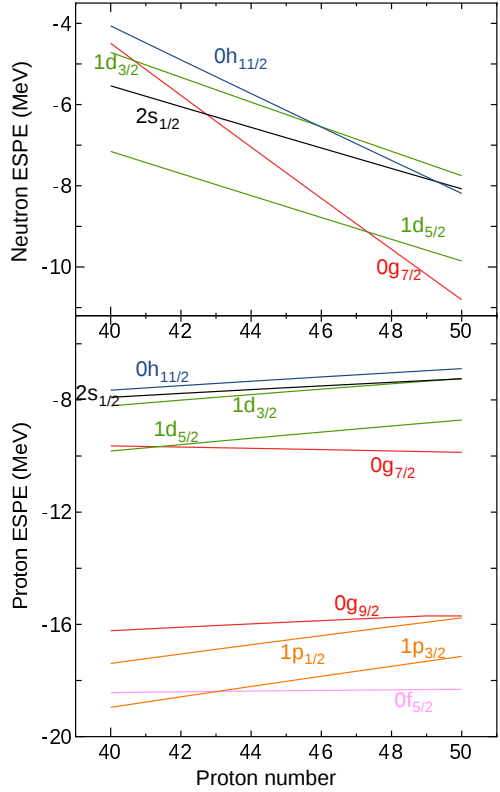}
  \caption{ Effective single-particle energies of the $N=82$ isotones
    for neutron orbits 
    (upper) and proton orbits (lower) as a function of the proton 
    number calculated with the Hamiltonian used in this study. 
  }
  \label{fig:ESPE_Zr-Sn}
\end{figure}

\begin{figure}[t]
  \centering
  \includegraphics[scale=0.4]{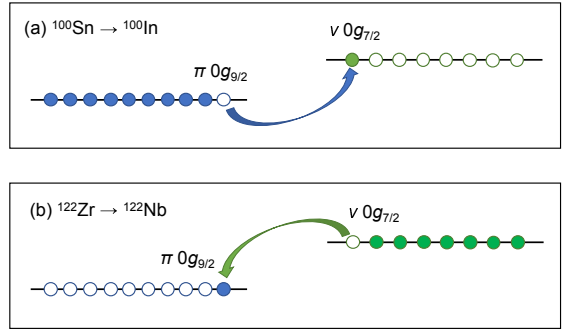}
  \caption{ Schematic illustration of the dominant single-particle 
transition in (a) the $\beta^+$ decay of $^{100}$Sn and 
(b) the $\beta^-$ decay of $^{122}$Zr. 
The filled and open circles denote particles and holes, respectively. 
  }
  \label{fig:gt_config}
\end{figure}

The $B(\textrm{GT})$ values predicted for $^{124}$Mo and $^{122}$Zr 
in the present study are the order of 
unity, although not reaching the measured value of $^{100}$Sn. Thus, they are 
new candidates for the superallowed Gamow-Teller transitions. 
Interestingly, those two regions of superallowed Gamow-Teller transition share 
the same underlying mechanism. 
In the extreme single-particle picture, the $\pi 0g_{9/2}$ orbit is completely 
filled and the $\nu 0g_{7/2}$ orbit is completely empty in $^{100}$Sn. 
Since the former and the latter orbits are the highest occupied and 
the lowest unoccupied ones, respectively, its low-energy Gamow-Teller transition 
is caused by the $\pi 0g_{9/2} \to \nu 0g_{7/2}$ transition. 
On the other hand, in $^{122}$Zr, the $\nu 0g_{7/2}$ orbit is completely 
filled and the $\pi 0g_{9/2}$ orbit is completely empty. 
As for the order of single-particle levels, 
Fig.~\ref{fig:ESPE_Zr-Sn} shows the evolution of the effective single-particle 
energies of $N=82$ isotones as a function of $Z$. 
For protons, the $\pi 0g_{9/2}$ orbit keeps the lowest unoccupied orbit in this range. 
For neutrons, although the $\nu 0g_{7/2}$ orbit is the lowest at $Z=50$ 
among the five orbits of interest, it goes up higher with decreasing $Z$ to 
finally be the second highest at $Z=40$. 
This is caused by a particularly strong attractive monopole interaction 
between $\pi 0g_{9/2}$ and $\nu 0g_{7/2}$ due to a cooperative attraction 
of the central and the tensor forces \cite{vmu}. 
This sharp change of the $\nu 0g_{7/2}$ orbit in going from $Z=40$ to 50 
is established 
from the energy levels of $^{91}$Zr and $^{101}$Sn, 
as mentioned in  \cite{vmu}. In $^{122}$Zr,
the $\nu 0g_{7/2}$ orbit is thus close to the highest occupied level, 
making a low-energy Gamow-Teller state by the $\nu 0g_{7/2} \to \pi 0g_{9/2}$ transition. 
If one is restricted to the configuration most relevant to the 
low-energy Gamow-Teller transition, the final state of the $^{122}$Zr decay, 
$(\nu 0g_{7/2})^{-1}(\pi 0g_{9/2})^{+1}$, 
is the particle-hole conjugation of that of the $^{100}$Sn decay, 
$(\pi 0g_{9/2})^{-1}(\nu 0g_{7/2})^{+1}$. 
A schematic illustration of these configurations are given 
in Fig.~\ref{fig:gt_config}. 
Accordingly, 
the $B(\textrm{GT})$ values from the vacuum to these single-particle 
configurations, i.e., those of Fig.~\ref{fig:gt_config}(a) and (b), 
are identical. 

One of the important ingredients for making $B(\textrm{GT})$ large 
in those nuclei is that the $B(\textrm{GT})$ value 
obtained within the single configuration of Fig.~\ref{fig:gt_config} 
(a) [and (b)] is also large. 
To be more specific, let us compare two cases as the initial state, 
(i) $|(\pi 0g_{9/2})^{10};J=0\rangle$ and 
(ii) $|(\pi 0g_{9/2})^{2};J=0\rangle$, where one proton can move to 
the $\nu 0g_{7/2}$ orbit through the Gamow-Teller transition. 
The case (i) corresponds to Fig.~\ref{fig:gt_config}(a) and 
yields $B(\textrm{GT})$=17.78 (without the quenching factor), 
whereas the case (ii) gives $B(\textrm{GT})$=3.56. 
The ratio of these two $B(\textrm{GT})$ values, 10 to 2, 
is just that of the number of protons in the initial state. 
This proportionality is well understood by remembering the Ikeda 
sum rule. 

Although the $B(\textrm{GT})$ value in the extreme single-particle 
picture is as large as 17.78 for the configurations of 
Figs.~\ref{fig:gt_config} (a) and (b), it is reduced in reality 
by the quenching factor 
and fragmentation over other excited states. 
To minimize fragmentation, it is desirable to suppress the 
level density with the same $J^{\pi}$ near the state of 
interest. $^{100}$Sn and $^{122}$Zr are doubly-magic (or semi-magic) 
nuclei, thus having a favorable condition for that. 
Another important factor to affect level density is excitation energy. 
As presented in Figs.~\ref{fig:gt-N82}, \ref{fig:gt-N81-2} 
and \ref{fig:zr122}, the low-energy Gamow-Teller peak is located 
stably at around 2~MeV by changing $Z$. 
This excitation energy is low enough to isolate the peak, 
if one remembers that the superallowed Gamow-Teller state 
from the $^{100}$Sn decay is located at $\sim 3$~MeV. 
In the present calculations, we do not include neutron excitations  
beyond the $N=82$ shell gap. Since these excitations typically cost 
more than 4~MeV by estimating from 
the first excitation energy of $^{132}$Sn, they 
probably do not contribute much to fragmentation. 

\begin{figure}[t]
  \centering
  \includegraphics[scale=0.5]{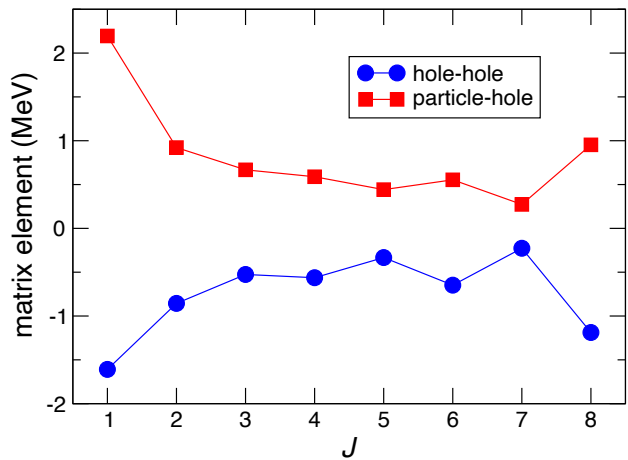}
  \caption{ Hamiltonian matrix elements concerning the $\pi 0g_{9/2}$ 
and $\nu 0g_{7/2}$ orbits used in this study. 
The circles and the squares are 
the hole-hole matrix elements,
$\langle (\pi 0g_{9/2})^{-1} (\nu 0g_{7/2})^{-1} | V | (\pi 0g_{9/2})^{-1} 
(\nu 0g_{7/2})^{-1} \rangle_J$ and the particle-hole matrix elements, 
$\langle \pi 0g_{9/2} (\nu 0g_{7/2})^{-1} | V | \pi 0g_{9/2} 
(\nu 0g_{7/2})^{-1} \rangle_J$, respectively. 
  }
  \label{fig:matele}
\end{figure}

One may wonder why the low-energy Gamow-Teller peak is kept at 
$E_x\sim 2$ MeV from $Z=48$ to $Z=40$ 
in spite of the sharp change of the $\nu 0g_{7/2}$ energy 
as shown in Fig.~\ref{fig:ESPE_Zr-Sn}. 
This is due to the nature of two-body Hamiltonian matrix elements. 
The low-energy Gamow-Teller state has always a neutron hole in $0g_{7/2}$. 
For the nuclei close to $Z=50$, this state has a few proton holes 
in $0g_{9/2}$, and thus its excitation energy is dominated by 
the hole-hole matrix element 
$\langle (\pi 0g_{9/2})^{-1} (\nu 0g_{7/2})^{-1} | V | (\pi 0g_{9/2})^{-1} 
(\nu 0g_{7/2})^{-1} \rangle_{J=1}$ as well as the single-particle energy of $\nu 0g_{7/2}$. 
As presented in Fig.~\ref{fig:matele}, this matrix element is 
the most attractive among the possible $J$ values. 
Hence the low-energy Gamow-Teller state is located lower 
than the simple estimate 
that the $0g_{7/2}$ orbit lies $\sim 3$ MeV below the Fermi surface 
at $Z=50$ (see Fig.~\ref{fig:ESPE_Zr-Sn}). 

This situation changes as more protons are removed from 
the $\pi 0g_{9/2}$ orbit. For the nuclei close to $Z=40$, 
the number of particles are smaller than the number of holes 
in the $0g_{9/2}$ orbit, and the particle-hole matrix element 
$\langle \pi 0g_{9/2} (\nu 0g_{7/2})^{-1} | V | \pi 0g_{9/2} 
(\nu 0g_{7/2})^{-1} \rangle_J$ plays a dominant role. 
In Fig.~\ref{fig:matele}, we also show the particle-hole matrix 
elements that are derived from the hole-hole matrix elements 
by using the Pandya transformation. 
The $J=1$ coupled matrix element has the largest positive value, 
thus losing 
the largest energy. 
This explains the calculated result 
that the low-energy Gamow-Teller state 
is not drastically lowered toward $Z=40$
as expected from the evolution of the $\nu 0g_{7/2}$ orbit, 
and also the observation that the corresponding state 
for the $^{100}$Sn decay is located 
at $\sim 3$~MeV \cite{hinke_Nature12}.  
It should be noted that this $J$ dependence is 
an example of the parabolic rule that holds for 
short-range attractive forces \cite{heyde_textbook}. 

To briefly summarize this section, the predicted superallowed 
Gamow-Teller transition toward $Z=40$ occurs due to 
(a) the full occupation of a neutron high-$j$ orbit ($\nu 0g_{7/2}$ 
in this case) 
and the emptiness of its proton spin-orbit partner 
($\pi 0g_{9/2}$ in this case) and 
(b) the low excitation energy of the $J=1$ particle-hole state 
created by these two orbits. 
Since the $J=1$ proton-neutron particle-hole matrix elements are generally 
most repulsive among possible $J$, 
it is needed to fulfill (b) that the $\nu 0g_{7/2}$ orbit 
and the $\pi 0g_{9/2}$ orbit 
are closed to the highest occupied orbit and the lowest unoccupied orbit,  
respectively. 
The tensor-force driven shell evolution plays a crucial role 
in satisfying this condition. 

\section{Summary}
\label{sec:summary}

We have constructed a shell-model effective interaction 
and performed large-scale shell-model calculations
of neutron-rich $N=82$ and $N=81$ nuclei
by utilizing our developed shell-model code
and the state-of-the-art supercomputers.
We demonstrated that the experimental binding
and excitation energies of neutron-rich $N=79,80,81$ nuclei
are well reproduced by
the available experimental data
including the low-lying excited states.
The present study gives the Gamow-Teller strength functions and
the $\beta$-decay half-lives of $N=82$ and $N=81$
nuclei, 
which are reasonably consistent with the available experimental data, 
and several predictions for further proton-deficient nuclei.
In these isotones, as the proton number decreases from $Z=49$ 
to $Z=42$, the proton $0g_{9/2}$ orbit becomes unoccupied and the
Gamow-Teller strengths of the low-lying states
increases because of the Pauli-blocking effect.
We predict that
the low-energy Gamow-Teller strength is further enlarged in $^{122}$Zr
to make its $\log ft$ value equivalent to that of the superallowed beta decay.
This is quite an analogous case to the so-called ``superallowed Gamow-Teller'' 
transition observed in $^{100}$Sn in terms of Gamow-Teller strength and
underlying mechanism. 

In the present work, we assume the contribution of the
first-forbidden transition is independent of the nuclides
and can be absorbed into 
a single quenching factor of the Gamow-Teller transition.
Further investigation to estimate
the first-forbidden decay especially for the $N=81$ isotones
is also expected.

\section*{Acknowledgment}

We acknowledge Takaharu Otsuka and Michio Honma for valuable discussions.
A major part of shell-model calculations was performed on CX400 
supercomputer of Nagoya University (hp160146)
and Oakforest-PACS supercomputer
(hp170230, hp160146, xg18i035).
The authors acknowledge valuable supports 
by ``Priority Issue on post-K computer''
(Elucidation of the Fundamental Laws and Evolution of the Universe)
and ``Program for Promoting Researches on the Supercomputer Fugaku''
(Simulation for basic science: from fundamental laws of particles
to creation of nuclei), MEXT, Japan.
We also acknowledge the KAKENHI grants (17K05433, 20K03981), JSPS, Japan.

\end{document}